\documentstyle[twoside,fleqn,espcrc2,epsf]{article}

\epsfverbosetrue

\newcommand{\AmS}{{\protect\the\textfont2
  A\kern-.1667em\lower.5ex\hbox{M}\kern-.125emS}}
\newcommand{\al}{\alpha}

\newcommand{\be}{\begin{equation}}
\newcommand{\ee}{\end{equation}}
\newcommand{\bea}{\begin{eqnarray}}
\newcommand{\eea}{\end{eqnarray}}
\newcommand{\bi}{\begin{itemize}}
\newcommand{\ei}{\end{itemize}}
\newcommand{\Dslash}{D\!\!\!\!\slash}

\newcommand{\NPB}[3]{{\em Nucl. Phys.} {\bf B{#1}} (19{#2}) {#3}}
\newcommand{\NPBproc}[3]{{\em Nucl. Phys.} {\bf B} (Proc. Suppl.)
           {\bf {#1}} (19{#2}) {#3}}

\newcommand{\PRD}[3]{{\em Phys. Rev.} {\bf D{#1}} (19{#2}) {#3}}

\newcommand{\etal}{{\em et al.}}

\title{A Study of PCAC for the Nonperturbative Improvement of the Wilson
	Action}

\author{Aida X. El-Khadra\\\vskip\baselineskip 
	Department of Physics,
	University of Illinois, 1110 W. Green St., Urbana, IL 61801-3080}

\begin{document}

\begin{abstract}
We present an exploratory study for the nonperturbative determination of the
coefficient of the ${\cal O}(a)$ improvement term to the Wilson
action, $c_{\rm SW}$. Following the work by L\"{u}scher {\it et al.},
we impose the PCAC relation as a nonperturbative improvement condition
on $c_{\rm SW}$, without, however, using the Schr\"{o}dinger functional
in our calculation.
\end{abstract}

\maketitle

\section{INTRODUCTION}

Much progress has been made in recent years in lattice QCD with the help
of improved actions. With better numerical tests, we also need better
determinations of the coefficients of the improvement terms.
We concentrate here on ${\cal O}(a)$ improvement for the fermion action,
and start with the Sheikholeslami-Wohlert
action \cite{sw},
\bea
S^{\rm lat} & = & a^4 \sum_x \left[\;\; m_0 \bar{\psi}\psi 
     + \bar{\psi}\Dslash^L \psi
     - \frac{1}{2}ar \bar{\psi} \Delta^L \psi \right. \nonumber \\
    & + & \left. 
   \frac{i}{4} c_{SW} \bar{\psi} \sigma_{\mu\nu} F_{\mu\nu}^L \psi
     \;\; \right] \;\;,
\eea
where $D^L_{\mu}$ and $\Delta^L$ are discretizations of the covariant
derivative and the laplacian, and $F_{\mu\nu}^L$ is 
the cloverleaf approximation to the $F_{\mu\nu}$ operator. 
At tree-level, $c_{SW} = 1$ \cite{sw}. With tadpole improvement,
\be
c_{SW} = \frac{1}{u_0^3} \;\;.
\ee
The one-loop correction is also known \cite{ww}:
\be \label{eq:1l}
c_{SW} = \frac{1}{u_0^3} \left[ 1 + 0.20 \al_V + {\cal O}(\al_V^2)
          \right] \;\;,
\ee
where $u_0$ is defined from the plaquette, 
$u_0 = \langle {\rm tr} U_{\Box} \rangle$. Finally, a recent 
nonperturbative determination \cite{alpha} parametrizes the improvement 
coefficient as:
\be
c_{SW} = \frac{1 - 0.656 g_0^2 - 0.152 g_0^4 - 0.054 g_0^6}{1-0.922 g_0^2}
       \;\;,
\ee
for $\beta \equiv 6/g_0^2 \geq 6.0$. This result has an unknown 
${\cal O}(a)$ error, which contributes an error of ${\cal O}(a^2)$ to
the action. It is therefore interesting to find alternative 
nonperturbative determinations of $c_{SW}$ to explore this 
uncertainty further.

\begin{figure}[t]
\begin{center}
\epsfxsize= 0.4\textwidth
\leavevmode
\epsfbox[17 303 487 595]{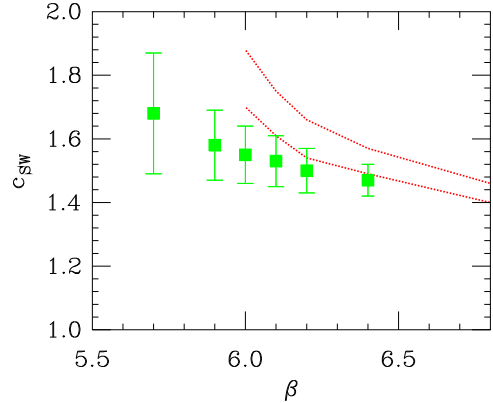}  
\end{center}
\caption[xxx]{$c_{SW}$ in comparison. $\Box$: one-loop result of 
Eq.~(\ref{eq:1l}), with $q^* \simeq 1/a$ and $\al_V$ as determined 
from the plaquette \cite{lm}; the error bars correspond to $\pm \al_V^2$.
--: nonperturbative $c_{SW}$ from Ref.~\cite{alpha}; the error band is taken
as $\pm 2 \,\%$ at $\beta=6.8$ and $\pm 5 \,\%$ at $\beta=6.0$. 
}
\label{fig:csw}
\end{figure}

Fig.~\ref{fig:csw} shows a comparison between the one-loop and the 
nonperturbative results for $c_{SW}$ as a function of $\beta$.
The comparison shows that the perturbative and nonperturbative results
for $c_{SW}$ are in reasonable agreement, once the uncertainties are
taken into account.

\section{THE PCAC RELATION}

The continuum PCAC relation, $\partial_{\mu} A_{\mu}(x) = 2 m\, P(x)$, is
modified on the lattice, due to lattice spacing artifacts. With
${\cal O}(a)$ improvement, the axial current is
\be \label{eq:ax}
   A_{\mu}^I(x) = A_{\mu}(x) + c_A \partial_{\mu} P(x) + {\cal O}(a^2)
\ee
where the improvement coefficient $c_A$ also needs to be determined.
The PCAC relation then is
\be \label{eq:pcac}
 \partial_{\mu} \langle A_{\mu}^I(x)\,\Pi(0) \rangle = 2 m \,
 \langle P(x) \, \Pi(0) \rangle + {\cal O}(a^2) \;\;,
\ee
This equation is valid for any (pion) operator, and can be used 
to find conditions that determine $c_{SW}$ and $c_A$, for example,
by varying the $\Pi$ operators, or the euclidean time dependence,
or the boundary conditions.

The Schr\"{o}dinger functional in Ref.~\cite{alpha} provides a
nice condition for $c_{SW}$. In this case, the boundary conditions 
induce a color background field, so that the vertex, 
$\bar{\psi} \sigma_{\mu\nu} F_{\mu\nu} \psi$, is sensitive to
the boundary conditions. 
We want to explore Eq.~(\ref{eq:pcac}) without using the
Schr\"{o}dinger functional in our numerical calculation, to make use
of existing lattices and shed light on the ${\cal O}(a)$ error 
associated with nonperturbative determinations of $c_{SW}$.

An alternative to the Schr\"odinger functional boundary conditions
is the momentum dependence of the two-point functions in 
Eq.~(\ref{eq:pcac}).
This can be seen as follows. The operator $\bar{\psi}\Dslash^{\;2}\psi$
is redundant at order ${\cal O}(a)$ \cite{sw}, and can be written
as
\be
   \bar{\psi} \Dslash^{\;2} \psi = \bar{\psi} D^2 \psi 
        - \frac{i}{2} \bar{\psi} \sigma_{\mu\nu}F_{\mu\nu} \psi
\ee
In the lattice action the first term is the Wilson term (a
discretized laplacian) and the omission of the second term from 
the Wilson action gives rise to the well known ${\cal O}(a)$ errors.
In the Sheikholeslami-Wohlert action the second term is added to
the lattice action, however, the $F_{\mu\nu}$ operator is discretized 
using plaquettes arranged in a ``clover-leaf''. 
Thus, the second term acquires a coefficient, $c_{SW}$, which is a 
function of $\al_s$ (and $m_0$).
Going to Fourier space, the momentum dependence is then sensitive
to $c_{SW}$. Thus, we can use operators with different momenta,
$\Pi(\vec{p})$ (or by translational invariance, 
$\partial_{\mu}A_{\mu}^I(\vec{p})$), in Eq.~(\ref{eq:pcac}) 
to determine $c_{SW}$.

Following Ref.~\cite{alpha}, we write Eq.~(\ref{eq:pcac}) as
\be  \label{eq:m}
   m(t) = \frac{1}{2} \,
          \frac{\partial f_{A_{\mu}}(t) + c_A \partial^2 f_P(t)}
               {f_P(t)} 
 = r(t) + c_A s(t) 
\ee
with
\be
f_{A_{\mu}}(t) = \langle A_{\mu}(t)\, \Pi(0) \rangle \;\;,\;
f_{P}(t) = \langle P(t)\, \Pi(0) \rangle \;\;.
\ee
Defining $m'(t)$ using a different operator $\Pi'$, it follows from 
Eq.~(\ref{eq:pcac}) that $m - m' = {\cal O}(a^2)$, 
if $c_{SW}$ and $c_A$ are tuned to their correct values. In the
combination
\be
M(t,t') = r(t) - s(t) \frac{r(t') - r'(t')}{s(t') - s'(t')}
\ee
$c_A$ drops out. Furthermore \cite{alpha}, $M = m + {\cal O}(a^2)$,
and
$\Delta M(t,t') = M(t,t') - M'(t,t') = {\cal O}(a^2)$, 
if $c_{SW}$ is tuned to its correct nonperturbative value.

\section{RESULTS}

The lattices used in this calculation were
originally generated for a different purpose. They were chosen for
this exploratory study for calculational convenience. In particular,
we have lattices with only two values of the improvement coefficient,
$c_{SW} = 0$ (Wilson case) and $c_{SW} \approx 1/u_0^3$.
At $\beta = 5.5$ ($8^3\times 16$) we have 500 (quenched) configurations
with one quark mass per $c_{SW}$ value.
At $\beta = 5.9$ ($16^3 \times 32$) we have 350 (quenched) configurations 
at $c_{SW} = 1.50$ only, but for a range of quark masses.

\begin{figure}[t]
\begin{center}
\epsfxsize= 0.45\textwidth
\leavevmode
\epsfbox[101 325 572 509]{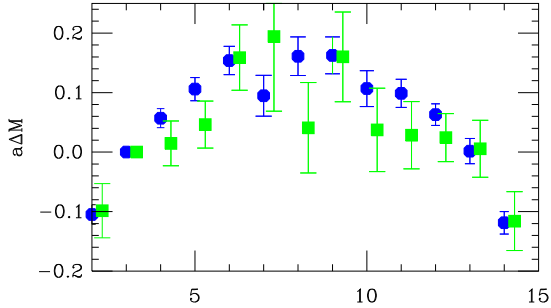}  
\end{center}
\caption[xxx]{$\Delta M(t,t'=3)$ vs. $t$ at $\beta = 5.5$. blue $\circ$:
$c_{SW} = 0$ and $\kappa = 0.169$, green $\Box$: $c_{SW} = 1.69$ and
$\kappa = 0.1423$. The squares are offset for readability.
}
\label{fig:dMts}
\end{figure}

Fig.~\ref{fig:dMts} shows a comparison of the time dependence of 
$\Delta M$ for two different values of $c_{SW}$ at $\beta = 5.5$. 
It appears that $\Delta M$ is different for the two values of $c_{SW}$
and that $\Delta M$ is closer to zero for $c_{SW} = 1.69$.
On both lattices, the quark mass roughly corresponds
to the strange quark mass, $\kappa \approx \kappa_s$. This large quark mass
was chosen, because at smaller quark masses the $\beta = 5.5$ lattices 
suffer from an increasing number of exceptional configurations \cite{ee}.

\begin{figure}[htb]
\begin{center}
\epsfxsize= 0.45\textwidth
\leavevmode
\epsfbox[91 309 579 594]{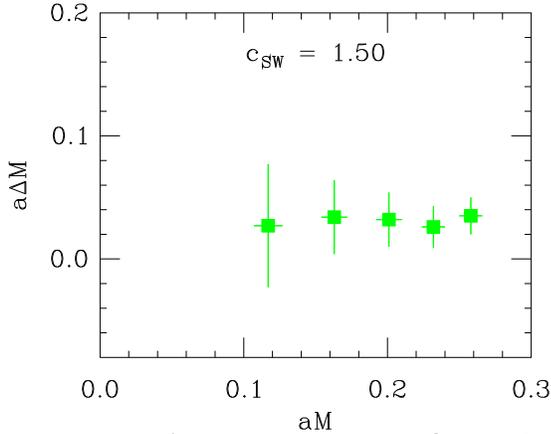}  
\end{center}
\caption[xxx]{$\Delta M$ vs. $aM$ at $\beta = 5.9$, $c_{SW} = 1.50$, 
and (from right to left) $\kappa = 0.1382, 0.1385, 0.1388, 0.1391, 0.1394$.
}
\label{fig:dMM}
\end{figure}

\begin{figure}[htb]
\begin{center}
\epsfxsize= 0.4\textwidth
\leavevmode
\epsfbox[ 69 302 549 587]{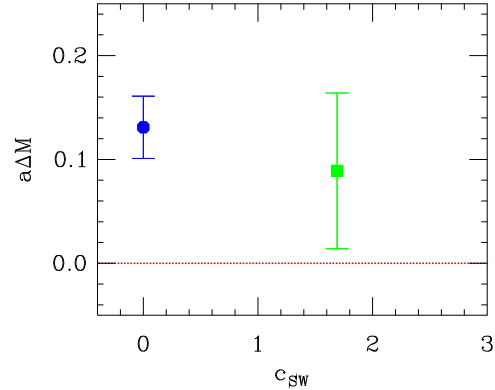}  
\end{center}
\caption[xxx]{$\Delta M$ vs. $c_{SW}$ at $\beta = 5.5$.
At $c_{SW} = 0$ $\kappa = 0.169$ and at $c_{SW} = 1.69$
$\kappa = 0.1423$.
}
\label{fig:dMcsw}
\end{figure}

We study the quark mass dependence of $\Delta M$ at $\beta = 5.9$, as
shown in Fig.~\ref{fig:dMM}. We find that the mass dependence is 
smaller than our statistical errors, consistent with previous results
\cite{alpha,scri}. The errors increase with decreasing quark mass.
Finally, Fig.~\ref{fig:dMcsw} shows our result for $\Delta M$ versus 
$c_{SW}$ at $\beta = 5.5$.

In conclusion, we showed that the momentum dependence appears to be a 
sensitive tool to determine $c_{SW}$ nonperturbatively. $\Delta M$
is independent of the light-quark mass, albeit within large statistical
errors. Our results indicate that the nonperturbative value for
$c_{SW}$ is larger than that suggested by tadpole improvement 
(with the plaquette), $c_{SW} > 1/u_0^3$. This is, of course,
consistent with the nonperturbative determinations of $c_{SW}$ based
on the Schr\"{o}dinger functional. Clearly, more work must be done
in order to use our method for a quantitative, nonperturbative 
determination of $c_{SW}$.

\section*{ACKNOWLEDGEMENTS}

I thank P. Lepage and M. L\"{u}scher for discussions, and of course
my collaborators, Andreas Kronfeld, Paul Mackenzie, Sin\'{e}ad Ryan, and 
Jim Simone. The numerical work presented here was performed on the 
Fermilab ACPMAPS computer. 
This work is supported in part through the DOE OJI program under 
grant no. DE-FG02-91ER40677 and by a fellowship from the Alfred P. Sloan 
foundation.
Finally, I thank the organizers for an enjoyable conference and especially
Ken Bowler and Richard Kenway for excellent advice on hillwalking 
in the Scottish Highlands.

\end{document}